\documentclass[11pt]{article}
\usepackage{eacl2014}
\usepackage{times}
\usepackage{url}
\usepackage{latexsym}
\usepackage{graphicx}
\usepackage{float}
\usepackage{amsmath}

\title{Optimizing Component Combination in a Multi-Indexing Paragraph Retrieval System}

\author{Boris Iolis \\
  Universit\'e Libre de Bruxelles (ULB) \\
  Boulevard du Triomphe \\
  Campus de la Plaine, CP212 \\
  B-1050 Brussels - Belgium \\
  {\tt biolis@ulb.ac.be} \\\And
  Gianluca Bontempi \\
  Universit\'e Libre de Bruxelles (ULB) \\
  Boulevard du Triomphe \\
  Campus de la Plaine, CP212 \\
  B-1050 Brussels - Belgium \\
  {\tt gbonte@ulb.ac.be} \\}

\date{}

\begin{document}
\maketitle
\begin{abstract}
We demonstrate a method to optimize the combination of distinct components in a paragraph retrieval system. Our system makes use of several indices, query generators and filters, each of them potentially contributing to the quality of the returned list of results. The components are combined with a weighed sum, and we optimize the weights using a heuristic optimization algorithm. This allows us to maximize the quality of our results, but also to determine which components are most valuable in our system. We evaluate our approach on the paragraph selection task of a Question Answering dataset.

\end{abstract}

\section{Introduction}

In the Information Retrieval domain, the combination of search results is a long studied problem, and can effectively increase the precision of the resulting system. As a result, many IR systems are designed to use multiple querying methods, and then combine the retrieved results, which is also called Data Fusion. Fox and Shaw \shortcite{fox} showed the effectiveness of combining multiple retrieval runs as opposed to selecting only one of them. Lee \shortcite{lee} combined search strategies using a simple, non-weighed sum. Vogt and Cotrell \shortcite{vogt} used a Linear Combination model, for which they optimized the weights to maximize the system's precision. Tiedemann \shortcite{tiedemann} employs a Genetic Algorithm to perform a similar optimization.

In this paper, we demonstrate a method to optimally combine the components of a paragraph retrieval system. Our approach is similar to those listed above, as we use a simple Linear Combination model: however, we include all our system's components in this model, and not only the querying modules. This allows us to not only optimize the mixture of querying methods, but also of filters and scorers.

We built our system using a multi-indexing architecture (several indices are being used for the same text corpus), and including some state-of-the-art query generators and filters. Our system retrieves a set of paragraphs from the text corpus based on the input query, which are then ranked according to their confidence scores and constitute the results list. All components are treated on the same level, and can equally contribute to the final confidence score associated to each retrieved paragraph. The components are combined using the Linear Combination model, and its weights are tuned using a Heuristic Optimization algorithm. Finally, we evaluate the results on a paragraph selection task using a Question Answering dataset.

The rest of the paper is organized as follows: in the next section, we present the architecture of our paragraph retrieval system. In Section \ref{sec:scoring}, we explain how all the components can be combined and tuned. Finally, Section \ref{sec:expresults} presents our experimental results, while Section \ref{sec:conclu} contains the conclusion and discussion on future work.

\section{System Architecture}
\label{sec:approach}

\begin{figure}[h]
\centering
\includegraphics[height=58mm,width=70mm]{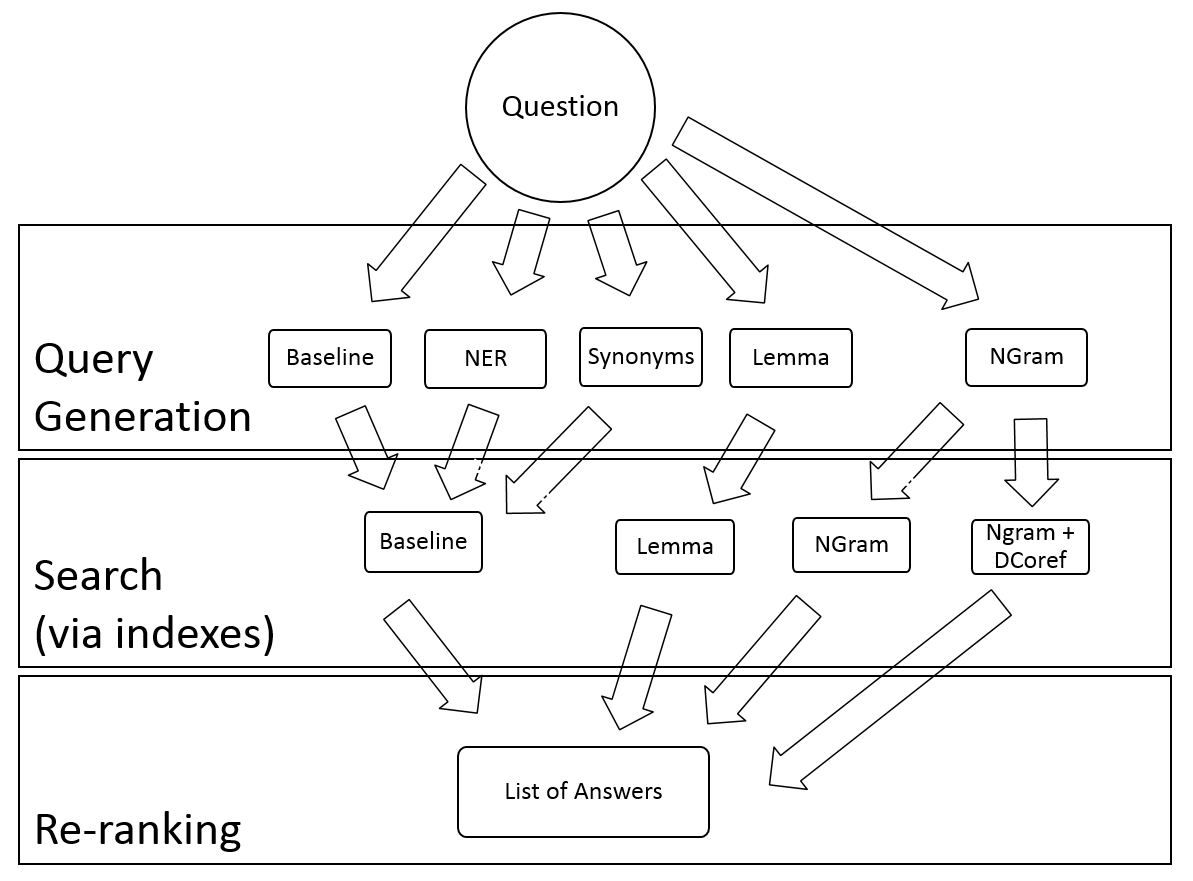}
\caption{Our multi-indexing search system architecture. }
\label{fig1}
\end{figure}

The architecture of our paragraph retrieval system is illustrated in Figure \ref{fig1}. It is based on the typical design of a Question Answering system (see for instance \cite{viewfromhere}), without answer extraction, as we try to retrieve a paragraph containing the correct answer to an input question instead of extracting the exact answer string from the text. We do however use multi-indexing, which is, to our knowledge, not so commonly studied in QA literature.

Starting from a single text corpus, we create a set of indices which will be used for querying. For each index, the text corpus is pre-processed in a distinct way. So far, our system implements the following four indices:

\begin{itemize}
\item {\bf Baseline}: standard inverted index on the text corpus, built using Lucene, which includes stopwords removal and simple tokenization. Only unigrams are indexed in this case
\item {\bf Lemmatization}: same as the baseline index but with a lemmatization step applied to the text corpus during pre-processing
\item {\bf NGrams}: same as the baseline index but with 2-grams and 3-grams added to the indexing terms
\item {\bf NGrams + Coreference Resolution}: same as NGrams, but with a coreference resolution step before indexing
\end{itemize}

We chose this multi-indexing approach in order to maximize the probability of retrieving the right paragraph in the querying stage (through at least one index). Typically, indexing in any specific way has its pros and cons; either we generalize too much (linking many similar terms to the same indexed term, for instance their common lemma), or not enough (indexing all words or ngrams separately). As queries will sometimes work better with more generalization, and sometimes with less, we are trying to get the best of both worlds by creating multiple indices and using them in parallel. Although there is a cost associated to creating and maintaining multiple indices, both in terms of disk space and pre-processing time, we believe that, even if the resulting improvements in recall are minimal, the benefits will outweigh the costs as long as the number of indices used is not excessively large.

In the querying stage, the input question is transformed into several queries, which is a common technique in IR and QA (see for instance \cite{dumais}). One query is generated for each index to match its specificities. For instance, to query the lemmatization index, the input question needs to undergo the same lemmatization step as did the text corpus. Furthermore, two additional query generation approaches are implemented, and both are used on the baseline index;

\begin{itemize}
\item {\bf Named Entity Recognition}: builds a query containing only the named entities found in the input question
\item {\bf Synonyms}: query expansion with synonyms based on WordNet \cite{wordnet}
\end{itemize}

Each query will return a list of paragraphs; in the last stage of our system, those paragraphs will be evaluated using a set of criteria, and then re-ranked in order to provide the most relevant list of paragraphs with regards to the original question. This re-ranking is based on our scoring framework, which is presented in the next section. The criteria we use at this stage are based on word counts (used extensively in IR and QA literature, for instance in \cite{ligozat}) and Latent Dirichlet Allocation (LDA) \cite{lda}.

\begin{itemize}
\label{evaluators}
\item {\bf Common words}: number of common words between the paragraph and the input question
\item {\bf Common 2-grams}: same as above but with 2-grams instead of single words
\item {\bf Common 3-grams}: same as above but with 3-grams
\item {\bf LDA-10}: cosine similarity between the probability vectors of the paragraph and the input question, based on a LDA model with 10 topics, trained on the text corpus
\item {\bf LDA-100}: same as above but with a 100 topics model
\end{itemize}

\section{Scoring Framework}
\label{sec:scoring}

\subsection{Computing the Scores}
\label{sec:computing}

Our method to score a paragraph is a simple application of the Linear Combination model to all the components of our system. We compile a list of criteria (we will call them features through the rest of the paper) consisting of all the query generators from the querying stage and the evaluators from the re-ranking stage. Each of those features gives a distinct score to each paragraph. For queries, the score of a paragraph is given by the Lucene confidence score if this paragraph was returned in the results list when using this query, and it is set to $0$ otherwise. For evaluators, this is straightforward. Each of these scores is then normalized using the Z-score normalization method \cite{zscore}. Finally, the overall score of a paragraph $c_i$ is computed as a linear combination of the features $f_j$, as shown below:

\begin{equation}
\label{linearcomb}
Score_{c_i} = \sum_{j=1}^{N} w_j \times f_{j}(c_i)
\end{equation}

where $N$ is the number of components (evaluators and query generators) of the system (in our case $N=11$); $f_{j}(c_i)$ is the score given by component $j$ to paragraph $i$; and $w_j$ are weights such that $\sum_{j=1}^{N} w_j = 1$

The actual ranking of the paragraph can be done by simply sorting them according to their score. This approach allows us to easily combine all the components of our system to obtain a global score for each paragraph.

\subsection{Tuning the Weights}
\label{sec:tuning}

In (\ref{linearcomb}), the weights should be tuned to maximize precision. They could be defined manually according to the quality of each feature (how relevant are the scores given by the feature), but unfortunately we do not have this knowledge beforehand. Also, evaluating each feature individually does not account for their diversity and complementarity when combined. Therefore, we decided to treat the tuning of those weights as a multivariate optimization problem, where the objective is to find the set of weights $w_j$ maximizing the overall performance of the system, according to an evaluation metric of interest. Though the cost function is not differentiable, we can still apply a wide variety of heuristic optimization methods (coordinate ascent, simulated annealing,...) to find the (approximate) best set of weights. For this work, we used a Differential Evolution algorithm \cite{storn} to perform this task, as it would allow us to demonstrate the effectiveness of our approach while being relatively simple to implement.

\section{Experimental Results}
\label{sec:expresults}

\subsection{Dataset}
\label{dataset}

We used the dataset from the ResPubliQA 2010 competition \cite{respubliqa}, containing a text corpus of 10,700 European parliament transcripts (taken from the JRC-Acquis\footnote{http://ipsc.jrc.ec.europa.eu/index.php?id=198} and Europarl\footnote{http://www.statmt.org/europarl/} collections), accompanied with a set of 200 questions, each having the correct answer provided (gold standard). The text documents are structured in numbered paragraphs of a few sentences each. We focused on the paragraph selection task (finding the paragraph containing the correct answer), which made it possible to perform automated assessment, by comparing the identifiers of the retrieved paragraphs to the gold standard. We compared our results with the work of \cite{molino}, who perform the same paragraph selection task on the same dataset.

\subsection{Results}
\label{sec:results}

Table \ref{table1} shows the results obtained by our system, first with all components combined in a naive way (all weights $w_j$ from (\ref{linearcomb}) being equal), and then with weight tuning as described in Section \ref{sec:tuning}. Our metric of choice is the Mean Reciprocal Rank (MRR), which gives a score of $1/r$ for each question, where $r$ is the position of the paragraph containing the right answer in the results list. For the weight tuning experiment, we used $20$ rounds of cross-validation to avoid over-fitting. In each round, the tuning was done on $190$ questions, and then evaluated on the remaining $10$. The result shown in the table is the average of those $20$ MRR scores.

\begin{table}[h]
\begin{center}
\begin{tabular}{l r}
\hline \noalign{\smallskip} 
\bf System & \bf MRR \\ \hline
Our system & 0.513 \\
Our system (with weight tuning) & 0.543 \\
QuestionCube (baseline) & 0.549 \\
QuestionCube (best) & 0.637 \\
\hline
\end{tabular}
\end{center}
\caption{\label{table1} Evaluation of our system, with and without weight tuning, and comparison with the QuestionCube system from \cite{molino}. }
\end{table}

We see that our system performs better when the combination of components is tuned with the Differential Evolution algorithm. Furthermore, our results are not so far behind the QuestionCube system. We are in line with the performance of their baseline version (which is already a full-fledged QA system on its own), but are behind the improved version from \cite{molino}, which uses far more advanced distributional semantic models than our simple LDA evaluators. 

Finally, our parameter tuning experiment gives us some insights on the added value of each component in our system; if a component is consistently given a weight of $0$ by the optimization algorithm, we can conclude it is not very valuable for the overall performance of our system. The average weight for each component across the $20$ cross-validation runs are shown in Table \ref{table2}. We can see that more than $90\%$ of the total weight was concentrated among three specific features: Lemmatization, Synonyms, and the 2-Grams evaluator. This preference for a very limited subset of components might suggest that our choice of components to implement might not have been the best, or that some of them might require additional fine-tuning.

\begin{table}
\begin{center}
\begin{tabular}{l r}
\hline \noalign{\smallskip} 
\bf Component  & \bf Average Weight \\ \hline
Baseline &  0.0 \\
Lemmatization &  0.303 \\
NGrams & 0.0 \\
NGram + Coref. & 0.033  \\
Named Entities & 0.0 \\
Synonyms & 0.357  \\
Common Unigrams & 0.0  \\
Common 2-Grams & 0.253  \\
Common 3-Grams & 0.054  \\
LDA-10 & 0.0 \\
LDA-100 & 0.0 \\
\hline
\end{tabular}
\end{center}
\caption{\label{table2} Average weights given to each component across the $20$ cross-validation runs of the Differential Evolution algorithm }
\end{table}

\section{Conclusion}
\label{sec:conclu}

In this paper, we demonstrated our method to efficiently combine the components of a paragraph retrieval system. We showed that using a heuristic optimization algorithm to tune this combination had a positive effect on the performance of our system. The overall performance is also in line with previous evaluations on the same dataset. Finally, we showed how this methodology could be used to evaluate the added value of each component which could be useful in our future work.

Now that we have this framework as a backbone, we can easily add new components to the system to make it more competitive in the future, as only the basic components have been integrated so far. As was shown in the weight tuning experiment, some effort may be required to understand why some of our components do not bring so much added value, and modify them to address this situation. Different optimization methods could also be implemented.

\section*{Acknowledgments}

We would like to thank David Verborgh for providing valuable input during the development of our application. This work is part of a PhD project funded by the Innoviris institute, via their Doctiris program, and carried out in cooperation with Mentis\footnote{http://www.mentis-consulting.be/}.

% Otherwise you can include your references as follows:
%% \begin{thebibliography}{}

%% \bibitem[\protect\citename{Aho and Ullman}1972]{Aho:72}
%% Alfred~V. Aho and Jeffrey~D. Ullman.
%% \newblock 1972.
%% \newblock {\em The Theory of Parsing, Translation and Compiling}, volume~1.
%% \newblock Prentice-{Hall}, Englewood Cliffs, NJ.

%% \bibitem[\protect\citename{{American Psychological Association}}1983]{APA:83}
%% {American Psychological Association}.
%% \newblock 1983.
%% \newblock {\em Publications Manual}.
%% \newblock American Psychological Association, Washington, DC.

%% \bibitem[\protect\citename{{Association for Computing Machinery}}1983]{ACM:83}
%% {Association for Computing Machinery}.
%% \newblock 1983.
%% \newblock {\em Computing Reviews}, 24(11):503--512.

%% \bibitem[\protect\citename{Chandra \bgroup et al.\egroup }1981]{Chandra:81}
%% Ashok~K. Chandra, Dexter~C. Kozen, and Larry~J. Stockmeyer.
%% \newblock 1981.
%% \newblock Alternation.
%% \newblock {\em Journal of the Association for Computing Machinery},
%%   28(1):114--133.

%% \bibitem[\protect\citename{Gusfield}1997]{Gusfield:97}
%% Dan Gusfield.
%% \newblock 1997.
%% \newblock {\em Algorithms on Strings, Trees and Sequences}.
%% \newblock Cambridge University Press, Cambridge, UK.

%% \end{thebibliography}

\end{document}